\documentclass[prd,a4paper,twocolumn,nofootinbib,preprintnumbers]{revtex4}
\usepackage[a4paper, hdivide={1.91cm,,1.165cm}, vdivide={1.83cm,,1.9cm}]{geometry}

\usepackage{amstext,amssymb}
\usepackage[intlimits]{amsmath}
\usepackage[english]{babel}
\usepackage[ansinew]{inputenc}
\usepackage{graphicx}
\usepackage{units}
\usepackage{color}
\usepackage{hyperref}

\hypersetup{
    pdftitle={Unbroken \texorpdfstring{$B-L$}{B-L} Symmetry},
    pdfauthor={Julian Heeck}
}

\begin{document}

\let \Lold \L
\def \L {\mathcal{L}} 
\let \epsilonold \epsilon
\def \epsilon {\varepsilon} 
\let \arrowvec \vec
\def \vec#1{{\boldsymbol{#1}}}
\newcommand{\dd}{\mathrm{d}}
\newcommand{\matrixx}[1]{\begin{pmatrix} #1 \end{pmatrix}} 
\newcommand{\BR}{\mathrm{BR}}
\newcommand{\diag}{\mathrm{diag}}

\allowdisplaybreaks

\title{Unbroken \texorpdfstring{$B-L$}{B-L} Symmetry}

\author{Julian \surname{Heeck}}
\email{julian.heeck@mpi-hd.mpg.de}
\affiliation{Max-Planck-Institut f\"ur Kernphysik, Saupfercheckweg 1, 69117 Heidelberg, Germany}
\affiliation{Institut f\"ur Theoretische Physik, Universit\"at Heidelberg, Philosophenweg 16, 69120 Heidelberg, Germany}

\preprint{\href{http://dx.doi.org/10.1016/j.physletb.2014.10.067}{Phys.\ Lett.\ B {\bf 739}, 256--262 (2014)}}
\preprint{\href{http://arxiv.org/abs/1408.6845}{arXiv:1408.6845}}

\begin{abstract}

The difference between baryon number $B$ and lepton number $L$ is the only anomaly-free global symmetry of the Standard Model, easily promoted to a local symmetry by introducing three right-handed neutrinos, which automatically make neutrinos massive. 
The non-observation of any $(B-L)$-violating processes leads us to scrutinize the case of unbroken gauged $B-L$;
besides Dirac neutrinos, the model contains only three parameters, the gauge coupling strength~$g'$, the St\"uckelberg mass~$M_{Z'}$, and the kinetic mixing angle~$\chi$. The new force could manifest itself at any scale, and we collect and derive bounds on $g'$ over the entire testable range $M_{Z'} = 0$--$\unit[10^{13}]{eV}$, also of interest for the more popular case of spontaneously broken $B-L$ or other new light forces. We show in particular that successful Big Bang nucleosynthesis provides strong bounds for masses $\unit[10]{eV}<M_{Z'}<\unit[10]{GeV}$ due to resonant enhancement of the rate $\overline{f}f\leftrightarrow \overline{\nu}_R\nu_R$. The strongest limits typically arise from astrophysics and colliders, probing scales $M_{Z'}/g'$ from TeV up to $\unit[10^{10}]{GeV}$.

\end{abstract}

\maketitle


\section{Introduction}

The Standard Model (SM) is a very successful description of particle physics, connecting the weak, strong, and electromagnetic forces to the framework of gauge symmetries. Only the latter two are unbroken symmetries, i.e.~bring with them conserved quantum numbers, color and electric charge, respectively.\footnote{Conserved charges follow from the residual unbroken \emph{global} symmetries of the unphysical \emph{local} gauge symmetries.} The search for additional forces has always driven particle physics, but has so far not been successful, resulting in tight bounds either on the strength or range of the new force. Today's research focuses almost exclusively on \emph{spontaneously broken} new gauge symmetries, mimicking the success of the electroweak theory. Still, the most exciting possibility would be an \emph{unbroken} new symmetry, and a corresponding conserved charge. In this letter, we point out that this possibility is not excluded as of now, if the conserved charge is the difference of baryon number $B$ and lepton number $L$.

Baryon and lepton numbers are the only classically conserved quantities in the SM, taking into account that the lepton mixing pattern observed in neutrino oscillations proves the non-conservation of the individual lepton numbers $L_{e,\mu,\tau}$ or linear combinations thereof. Ignoring new forces that do not act on SM particles, e.g.~unbroken hidden forces in the dark matter sector, it is clear that the only new exact symmetry can be a linear combination of $B$ and $L$. Classical symmetries of the Lagrangian can be broken at quantum level through triangle anomalies, which have to be canceled in order to obtain a valid quantum field theory. For $B-L$, the anomalies can be canceled simply by introducing three SM-singlet right-handed neutrinos $\nu_R$, which automatically lead to massive active (Dirac) neutrinos---a very welcome side effect. Gauging any other linear combination $X$ of $B$ and $L$ would require the introduction of chiral fermions charged under $SU(2)_L\times U(1)_Y$ and is by now viable only if $X$ is spontaneously broken to generate fermion masses above (and disconnected from) the electroweak scale (\emph{highly} fine-tuned models not withstanding). 

$U(1)_{B-L}$ hence emerges as the only possible unbroken gauge symmetry acting on known particles beyond electromagnetism and color. This is further corroborated by the fact that we have yet to observe any $B-L$ violating process despite decade-long searches, most prominently the search for $\Delta (B-L) = 2$ neutrinoless double beta decay~\cite{Rodejohann:2011mu}. Even the baryon asymmetry of our Universe is no argument for breaking $B-L$, because the Dirac nature of neutrinos gives rise to an elegant leptogenesis mechanism under the name of neutrinogenesis even for conserved $B-L$~\cite{Dick:1999je}.

It is therefore worthwhile to study the implications of an unbroken $U(1)_{B-L}$ symmetry as an alternative point of view to direct searches for $B-L$ violation.
This scenario has rarely~\cite{Feldman:2011ms, Feng:2013zda} been considered; new (fifth) forces coupled to baryon number~\cite{Lee:1955vk}, lepton number~\cite{Okun:1969ey}, or $B-L$~\cite{Carlson:1986cu} have of course been studied before, but never as \emph{unbroken}, and hence never with \emph{Dirac} neutrinos. This will be of great importance in the following, as the coupling to light right-handed neutrinos severely constrains the new force.

We briefly present the key aspects of the model in Sec.~\ref{sec:model} and then go on to collect, update, and improve bounds on the $U(1)_{B-L}$ coupling strength $g'$ for $Z'$ masses up to 10 TeV in Sec.~\ref{sec:limits}, presented in Fig.~\ref{fig:limits}. We comment on the applicability of these limits to other closely related models in Sec.~\ref{sec:comments} and finally conclude in Sec.~\ref{sec:conclusion}.

\section{Model}
\label{sec:model}

Let us briefly review the key features of this simple model~\cite{Feldman:2011ms}.
A local $U(1)_{B-L}$ requires the addition of three right-handed neutrinos to the SM in order to cancel gauge anomalies. For unbroken $B-L$, these states form Dirac particles with the active left-handed neutrinos via Yukawa couplings to the Higgs doublet $y_{\alpha \beta} \overline{L}_{\alpha} H \nu_{R,\beta}$, resulting in the neutrino mass matrix 
\begin{align}
M_{\alpha \beta} = y_{\alpha \beta} |\langle H\rangle | = U \diag (m_1, m_2, m_3) V_R^\dagger
\end{align}
after electroweak symmetry breaking.
Here, $U$ denotes the unitary leptonic mixing matrix relevant for charged-current interactions (in the standard parametrization with vanishing Majorana phases~\cite{Beringer:1900zz}), $m_i$ are the neutrino masses, and $V_R$ is an unphysical unitary matrix. 
Current values for the neutrino parameters can be found in Ref.~\cite{GonzalezGarcia:2012sz}. 
Note that experimental upper bounds on the neutrino masses of order eV imply very small values for the Yukawa couplings $|y_{\alpha \beta}|\lesssim 10^{-11}$ but pose no real problem to the model.

Dirac neutrinos aside, an unbroken $U(1)_{B-L}$ brings with it only one more particle: the gauge boson $Z'$, coupled to the $B-L$ current $j_{B-L}\equiv j_B - j_L$ via $g' Z'_\mu j^\mu_{B-L}$. All fermions---including neutrinos---are described by Dirac fermions after electroweak symmetry breaking, leading to vector-like $Z'$ couplings to these mass eigenstates:
\begin{align}
 \L\,\supset \, g' Z'_\mu \sum_{\mathrm{fam}} \left[ \tfrac{1}{3} \left(  \overline{u} \gamma^\mu u+ \overline{d} \gamma^\mu d \right) - \overline{e} \gamma^\mu e - \overline{\nu} \gamma^\mu \nu \right] .
\label{eq:vectorcoupling}
\end{align}
There are in particular no flavor changing neutral currents.
The $Z'$ itself is allowed to have a mass $M_{Z'}$ via the St\"uckelberg mechanism~\cite{Stueckelberg:1900zz} without breaking the (global) $B-L$ symmetry~\cite{Feldman:2011ms}.
This mass is a free parameter of the model, not connected to a vacuum expectation value; it is notably disconnected from, e.g., neutrino masses, and there exists no theoretically preferred value.\footnote{Note though that small values for both $g'$ and $M_{Z'}$ are technically natural in the sense of 't~Hooft~\cite{tHooft:1979bh} in that all radiative corrections are again proportional to $g'$ and $M_{Z'}$, respectively.}

Lastly, the $Z'$ can kinetically mix with the hypercharge boson through a coupling $\frac12\sin \chi F^Y_{\mu\nu}F'^{\mu\nu}$~\cite{Galison:1983pa, Holdom:1985ag}, effectively coupling the $Z'$ to the hypercharge current. We will neglect this kinetic mixing angle $\chi$ in the following for simplicity, but comment on its effects in Sec.~\ref{sec:comments}.

Let us make a couple of remarks to compare unbroken $B-L$ with the better-studied case of spontaneously broken $B-L$ (``Majorana $B-L$''), featuring three heavy and three light Majorana neutrinos.
The decay width into fermions is given by
\begin{align}
\begin{split}
 \Gamma (Z'\to \overline{f} f) = &\tfrac{1}{3} \alpha' M_{Z'} \left( 1 + 2\frac{m_f^2}{M_{Z'}^2}\right) \sqrt{1 - 4\frac{m_f^2}{M_{Z'}^2}} \\
&\times 
\begin{cases}
 1, & f = \text{lepton},\\
 1/3, & f = \text{quark},
\end{cases}
\end{split}
\label{eq:decayrate}
\end{align}
with the $B-L$ fine-structure constant $\alpha' \equiv g'^2/4\pi$. 
The invisible width of our $Z'$ is then governed by the decay into the three light Dirac neutrinos $\nu = \nu_L + \nu_R$:
\begin{align}
 \Gamma_\mathrm{inv} (Z') = 3\times \Gamma (Z'\to \overline{\nu} \nu) = \alpha' M_{Z'} \,,
\label{eq:invisible_Dirac_width}
\end{align}
which effectively counts the number of light neutrinos, in complete analogy to the invisible width of the $Z$, which however only counts the number of light \emph{left-handed} neutrinos. The invisible $Z'$ width of unbroken $B-L$ is hence up to three times larger than in Majorana $B-L$ (depending on the mass of the heavy neutrinos), due to the additional decay into $\nu_R$. This is typically irrelevant for $Z'$ collider searches, but will be very relevant for lighter $Z'$, and would of course be of great interest in case a $Z'$ resonance is found.

\begin{figure}
\centering
\includegraphics[width=0.45\textwidth]{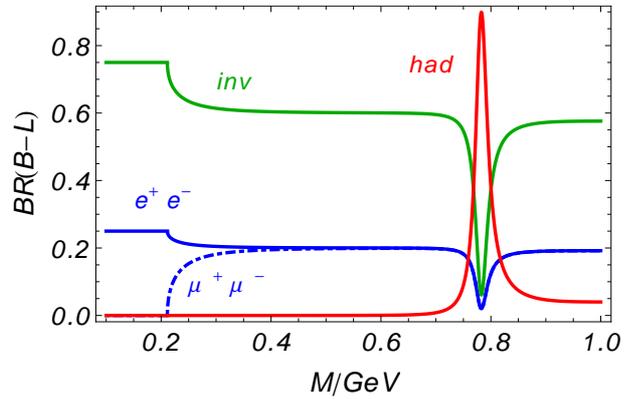}
\caption{ $Z'$ branching ratios to $e^+ e^-$ (blue), $\mu^+\mu^-$ (blue, dotdashed), neutrinos (green), and hadrons (red).}
\label{fig:branchingratios}
\end{figure}

For $Z'$ masses around $\Lambda_\mathrm{QCD}\sim \unit[200]{MeV}$, the hadronic $Z'$ decay is in principle more involved than Eq.~\eqref{eq:decayrate} because of hadronization. In the vector-meson dominance approach, $Z'$ shares the quantum numbers of the $\omega$ meson, because it carries no isospin. It therefore has similar decay modes~\cite{Tulin:2014tya}, most importantly with highly suppressed $Z'\to \pi\pi$ rate.
Hadronic decays are still possible for $M_{Z'} > m_\pi$, with the dominating channel $Z'\to \pi^0 \gamma$ below $\unit[600]{MeV}$ and partial width~\cite{Tulin:2014tya}
\begin{align}
 \Gamma (Z'\to \pi^0 \gamma) = \frac{\alpha' \alpha_\mathrm{EM}}{96\pi^3 f_\pi^2}  M_{Z'}^3 \left( 1 -\frac{m_\pi^2}{M_{Z'}^2}\right)^3 |F_\omega (M_{Z'}^2)|^2 . 
\end{align}
Here $F_\omega (s) \simeq (1-s/m_\omega^2- i \Gamma_\omega/m_\omega)^{-1}$ denotes the Breit--Wigner form factor of the $\omega$ meson.
For $\unit[600]{MeV} \lesssim M_{Z'} \lesssim \unit[1]{GeV}$, the dominant hadronic decay is $Z'\to \pi^+ \pi^- \pi^0$, with a rather complicated analytical expression~\cite{Tulin:2014tya}. (Here, we also include the $\rho$ width in the propagators in order to cross the $Z'\to \rho \pi$ threshold.) Above GeV, $s$-quark Kaon channels open up and the $Z'$ becomes $\phi$-like.
From Fig.~\ref{fig:branchingratios} we see that the hadronic branching ratio for $M_{Z'}< \unit{GeV}$ is completely negligible away from the $\omega$ resonance. This is in stark contrast to the hadronic decays of hidden photons~\cite{Batell:2009yf}, which feel the much broader $\rho$ resonance. The reason for this is the isospin-violating coupling of hidden photons to electric charge, allowing them to easily mix with both the $\omega$ (isospin 0) and the $\rho$ (isospin 1) meson.

Regarding the baryon asymmetry of our Universe, let us briefly sketch the $B-L$ conserving \emph{neutrinogenesis} mechanism, following Ref.~\cite{Dick:1999je}: here it is crucial to note that the small Yukawa couplings $|y_{\alpha \beta}|\lesssim 10^{-11}$ of our Dirac neutrinos are insufficient to put the $\nu_R$ into thermal equilibrium in the early Universe. The goal is then to create a lepton asymmetry $\Delta_{\nu_R}$ in the $\nu_R$ sector that is formally canceled by an opposite asymmetry of the left-handed leptons $\Delta_{\nu_L} = - \Delta_{\nu_R}$, which can be achieved with the introduction of e.g.~two very heavy unstable scalar doublets. Since only the left-handed leptons are in thermal equilibrium with the SM plasma, a baryon asymmetry $\Delta_B$ will be generated by the sphalerons using $\Delta_{\nu_L}$ in typical leptogenesis manner. In this letter we will not be explicitly concerned with neutrinogenesis, so we ignore the additional scalars and constraints. Let us note though that the newly introduced gauge interactions via $Z'$ do not invalidate this mechanism, because they conserve the individual particle numbers and hence do not erase $\Delta_{\nu_R}$ or $\Delta_{\nu_L}$; rapid gauge interactions will of course put the $\nu_R$ in equilibrium and hence increase the number of relativistic degrees of freedom, to be discussed in Sec.~\ref{sec:limits}.

Note that the gauge boson $Z'$ can in principle also be stable enough to be dark matter. For this, we have to make sure its lifetime $\tau$ surpasses that of our Universe, which requires
\begin{align}
 \tau = \frac{1}{\Gamma} \propto \frac{1}{\alpha' M_{Z'}} > \tau_\mathrm{Universe} \simeq \frac{1}{\unit[10^{-32}]{eV}}\,,
\label{eq:stability}
\end{align}
the dominant decay mode for light $Z'$ being $Z'\to \overline{\nu}\nu$ (unless $M_{Z'} < 2 m^\nu_\mathrm{lightest}$, then $Z'\to 3\gamma$ is the only channel). Using a misalignment mechanism in complete analogy to the hidden photon case~\cite{Nelson:2011sf} (see also Ref.~\cite{Arias:2012az}) could then make the $Z'$ a \emph{cold} dark matter candidate. We will not discuss this scenario any further in this letter.

\section{Limits}
\label{sec:limits}

The main signature of unbroken $B-L$, beyond Dirac neutrinos, is the new gauge boson $Z'$. As mentioned above, both coupling strength and mass are free parameters, disconnected from other observables. One can therefore consider $(g',M_{Z'})$ in the range $[0,1]\times [0,\unit[10]{TeV}]$ with equal motivation. While still a lamppost search, it is a mighty big one.

We will now collect and update bounds on the parameters $g'$ and $M_{Z'}$, ultimately resulting in Fig.~\ref{fig:limits}. The first (and only) survey across all masses can be found in Ref.~\cite{Carlson:1986cu}, which we refine and improve. Note that we will set the kinetic mixing angle $\chi$ to zero in our analysis for simplicity, but comment on its effects later on in Sec.~\ref{sec:comments}. The mass range for $B-L$ below GeV was recently covered in Ref.~\cite{Harnik:2012ni}; the mass range above MeV in Ref.~\cite{Williams:2011qb} (including kinetic mixing). However, none of the above have specifically considered \emph{unbroken} $B-L$, which has the $Z'$ coupled to \emph{Dirac} neutrinos; this makes especially Big Bang nucleosynthesis (BBN) bounds far more important and deserves a discussion.

Many limits are obtained by translating well-known hidden photon limits, see Ref.~\cite{Essig:2013lka} for a recent review. The main differences between these models are the invisible width of $Z'_{B-L}$, which reduces the (typically relevant) branching ratio into electrons and muons, and the additional coupling to neutral particles (neutrons and neutrinos). Unfortunately the literature is inconsistent when it comes to the definition of the kinetic mixing angle of hidden photons, employing either mixing with the hypercharge or electromagnetic field strength tensor. The former represents the proper gauge invariant structure and reduces to the latter in the low-energy limit. The two definitions differ by the cosine of the Weinberg angle---$\chi_\text{EM} = \chi_Y \cos\theta_W$ for small $\chi$---and care has been taken below to ensure consistency in the resulting limits on $g'$.

Note that the limits on $g'$ and $M_{Z'}$ are directly applicable to \emph{any} $B-L$ gauge boson coupled to Dirac neutrinos, even if the $U(1)_{B-L}$ is broken by $n\neq 2$ units (employed in Refs.~\cite{Heeck:2013rpa, Heeck:2013vha}). They are also applicable to models where $B-L$ is broken by two units, but all six Majorana neutrinos are light, so that the invisible $Z'$ width does not change compared to the unbroken $B-L$ case discussed here. For Majorana $B-L$ with (some) heavy Majorana neutrinos, not all limits are applicable and we will comment on that in due time.

\subsection{Modified gravity and fifth force searches}

A distinct feature of the $B-L$ force, compared to hidden photons in particular, is the coupling to neutral particles. As such, a light $Z'$ mediates a force between astrophysical bodies that depends on their neutron number and thus violates the weak equivalence principle.\footnote{The $B-L$ contributions from electrons and protons cancel each other in electrically neutral objects.} Torsion-balance experiments~\cite{Wagner:2012ui, Adelberger:2009zz} set very strong limits on such long-range forces, which are most importantly also applicable for a completely massless $Z'$, resulting in
\begin{align}
 g' < 10^{-24} \, \text{ at } 95 \% \text{ C.L. for } M_{Z'} <\unit[10^{-14}]{eV}\,.
 \label{eq:long_range_B-L}
\end{align}
Distances $\lambda \equiv 1/M_{Z'}$ from cm to $\unit[10]{\mu m}$ are constrained by experiments that test the gravitational inverse square law $F\propto 1/r^2$, which would be modified in the presence of a light $Z'$~\cite{Adelberger:2009zz, Yang:2012zzb}.
At even smaller distances, it is the Casimir effect (or Van der Waals forces) that sets the strongest limits on the modified $1/r^2$ law~\cite{Bordag:2001qi, Decca:2005qz, Sushkov:2011zz}. Neutron scattering limits are inferior to astrophysical constraints in the mass range $\unit[1]{eV}< M_{Z'} <\unit[10^5]{eV}$, but offer the best laboratory limits~\cite{Pokotilovski:2006up, Nesvizhevsky:2007by}.

\subsection{Stellar evolution}

For $Z'$ masses between $\unit[0.1]{eV}$ and $\unit[0.1]{GeV}$, strong astrophysical limits arise from the additional energy loss provided by the $Z'$, either in the Sun, heavier stars, or red giants. 
These limits, dominant for $\unit[10]{eV} \lesssim M_{Z'} \lesssim \unit[100]{keV}$, can be readily translated from hidden-photon limits~\cite{Redondo:2013lna} via $g' \mathrel{\hat=} e \chi$, because the relevant coupling to electrons (and protons) has the same structure for electric charge and $B-L$. Minor modifications arise due to the $B-L$ coupling to neutrons inside the star, which we neglect here.

The stellar evolution is actually modified two-fold: a light $Z'$ can carry away energy, similar to hidden photons, but the $Z'$ also enables the production of the sterile $\nu_R$; millicharged-particle limits~\cite{Davidson:2000hf, Harnik:2012ni} then constrain roughly $g' \lesssim 10^{-14}$ using red giants, which set the strongest limits in the range $\unit[0.1]{eV} \lesssim M_{Z'} \lesssim \unit[10]{eV}$. White dwarf cooling via $\nu_R$ emission yields constraints in the MeV--GeV region~\cite{Dreiner:2013tja} of similar order as neutrino scattering experiments (next section). This white dwarf limit is not shown in Fig.~\ref{fig:limits} in order to avoid cluttering.

From the supernova SN1987A we get very strong limits on $Z'$ in the $\unit[100]{keV}$--$\unit[100]{MeV}$ range~\cite{Dent:2012mx}, again adopted from hidden photons. These limits do, however, neglect some important plasma effects and are expected to change upon reanalysis. We show this limit in dashed lines in Fig.~\ref{fig:limits} to remind the reader of the subtleties involved.
Once again, the fact that the $Z'$ couples to sterile $\nu_R$ yields an additional supernova cooling bound; we adopt the limits from Ref.~\cite{Dreiner:2013mua}, based on on-shell $Z'$ production in $e^+ e^-$ annihilations, using $\BR (Z' \to \nu_R\nu_R)$ from Sec.~\ref{sec:model}. For heavier $Z'$, off-shell $\nu_R$ production becomes dominant, and one can derive bounds on $M_{Z'}/g'$~\cite{Raffelt:1987yt, Barbieri:1988av, Grifols:1989qm} of order TeV (not shown in Fig.~\ref{fig:limits}).

The stellar evolution bounds are of course not as refined as laboratory limits and should be understood as estimates; there are no confidence levels attached to them. A dedicated analysis of the $Z'$ and $\nu_R$ interplay in stellar evolution is expected to improve on the given bounds, but lies beyond the scope of this letter.

\subsection{Neutrino scattering}

Scattering of solar neutrinos on electrons in Borexino~\cite{Bellini:2011rx} is sensitive to $Z'$ exchange, recently discussed in Refs.~\cite{Harnik:2012ni,Chiang:2012ww, Laha:2013xua} for an MeV vector boson coupled to left-handed neutrinos and charged leptons. For sub-MeV masses, Gemma gives stronger limits~\cite{Harnik:2012ni}, but still weaker than astrophysics. These limits surpass magnetic-moment limits, and make in particular a resolution of the muon's $g-2$ anomaly impossible within gauged $B-L$ (or gauged $U(1)_L$ for that matter~\cite{Lee:2014tba}). 
Neutrino--quark scattering also yields strong limits of order $M_{Z'}/g' \gtrsim \unit{TeV}$ for $Z'$ masses above $\unit[10]{GeV}$~\cite{Carlson:1986cu, Williams:2011qb}.
Precise measurements have been performed by NuTeV~\cite{Zeller:2001hh}, which may however suffer from systematic errors. We therefore take a very conservative bound of $M_{Z'}/g' > \unit[0.4]{TeV}$, following a recent reanalysis~\cite{Escrihuela:2011cf}, about a factor 3--4 weaker than previous limits~\cite{Davidson:2003ha, Williams:2011qb}.

\subsection{Beam dump}

We take the $95\%$~C.L. limits from hidden photons from Ref.~\cite{Andreas:2012mt}, containing data from E774, E141, Orsay, KEK, and E137.
The lower (horizontal) limit on the coupling can be translated simply via $g' \mathrel{\hat=} e \chi $, because the number of events for small coupling is given by~\cite{Bjorken:2009mm}
\begin{align}
 N \propto \frac{g_{Z' ee}^2 }{M_{Z'}} \,\Gamma (Z'\to e^+ e^-)\,,
\end{align}
\emph{independent} of the invisible $Z'$ width. The width \emph{is} on the other hand important for the upper limits (diagonal $\chi\sim 1/M_{Z'}$), which correspond to a fast $Z'$ decay (with Lorentz factor $\gamma = E/M_{Z'}$) inside the shielding of length $L_\mathrm{sh}$; here, the number of events goes with
\begin{align}
 N  \propto \frac{g_{Z' ee}^2}{M_{Z'}^2} \BR (Z'\to e^+ e^-) \, \exp\left( - L_\mathrm{sh} \frac{M_{Z'}}{E}  \Gamma_\mathrm{total}\right),
\end{align}
sensitive to $ \Gamma_\mathrm{total}$, and hence $\Gamma_\mathrm{inv}$, due to the dominating exponential factor. This shows that the upper limits have to be translated by comparing the \emph{total width}. Below the muon threshold, this simply corresponds to $g' \mathrel{\hat=} e \chi/2 $, because we have for $2 m_e\ll M_{Z'} < 2 m_\mu$:
\begin{align}
\Gamma_\mathrm{total}^\mathrm{HP} = \frac{\alpha \chi^2}{3} M_{Z'}\,, && \Gamma_\mathrm{total}^{B-L} = \frac{\alpha'}{3} M_{Z'} \left[3 (\nu) + 1 (e^-)\right] .
\end{align}
The excluded area is hence \emph{smaller} for unbroken $B-L$ than maybe expected, due to the additional decay channel into neutrinos.

\subsection{BaBar}

BaBar gives great limits on hidden photons in the range $\unit[20]{MeV}< M_{Z'} < \unit[10]{GeV}$~\cite{Lees:2014xha} (rescaled to $95\%$~C.L.) from the coupling to electrons in the process $e^+ e^-\to \gamma Z'$, $Z'\to e^+ e^-$ ($\mu^+\mu^-$). Our larger invisible width will reduce the bounds, similar to the beam dumps discussed above (by a factor of 2 below the muon threshold), so we translate
\begin{align}
 g'  \mathrel{\hat=} e \chi \sqrt{\frac{\BR (A_\mathrm{HP}'\to e^+ e^-)}{\BR (Z'\to e^+ e^-)}}
\end{align}
over the entire mass range. These are the strongest limits for masses $\unit[5]{GeV}< M_{Z'} < \unit[10]{GeV}$. 
The fixed-target experiments APEX~\cite{Abrahamyan:2011gv} and MAMI~\cite{Merkel:2014avp} provide similar, slightly weaker, constraints, not shown in Fig.~\ref{fig:limits}.

\subsection{Thermalization in the early Universe}
\label{sec:bbn}

Big Bang nucleosynthesis (BBN) describes successfully our Universe at temperatures around MeV, and places strong bounds on the number of relativistic degrees of freedom. The latter are typically parameterized via $N_\mathrm{eff}$, the effective number of neutrinos, predicted to be 3 by the SM.  We take $ N_\mathrm{eff} < 4$ as a conservative $95\%$~C.L.~limit from BBN~\cite{Mangano:2011ar}, which in particular forbids the thermalization of our three light right-handed neutrinos $\nu_R$, and thus constrains the strength of the $B-L$ gauge interactions that would put them in equilibrium. The interaction rate of $\nu_R$ therefore has to be smaller than the Hubble expansion rate $H(T) \sim T^2/M_\mathrm{Pl}$ around $T\sim \unit[1]{MeV}$. Such reasoning has long since been used to constrain right-handed neutrino interactions~\cite{Steigman:1979xp, Olive:1980wz}.

To calculate the thermally averaged interaction rate for $\overline{f} f\leftrightarrow \overline{\nu}_R\nu_R$ induced by $Z'$ exchange we follow Ref.~\cite{Barger:2003zh}:
\begin{align}
\begin{split}
  \langle \Gamma (\overline{f} f\leftrightarrow \overline{\nu}_R\nu_R) \rangle  &= \frac{2}{n_{\nu_R}(T)} \int \frac{\dd^3 \vec{p}}{(2\pi)^3}\frac{\dd^3 \vec{k}}{(2\pi)^3} \, f_\nu (p) f_\nu (k)\\
 &\quad \times  v_\mathrm{M} \sigma_{\overline{\nu}_R\nu_R \to \overline{f} f} (s)\,,
\end{split}
\end{align}
with the Fermi--Dirac distribution $f_\nu (k) = (e^{k/T}+1)^{-1}$, the $\nu_R$ number density $n_{\nu_R} = 3 \zeta (3) T^3/2 \pi^2$, and the M{\o}ller velocity $v_\mathrm{M}$. The interaction cross section $\sigma$ is to be evaluated at the center-of-mass energy $s = 2 p k (1-\cos\theta)$, $\theta$~being the angle between the two colliding $\nu_R$ particles. Unlike Ref.~\cite{Barger:2003zh}, we do not restrict ourselves to the limiting case $M_{Z'} \gg T$, but rather use the full $Z'$ propagator. For $M_{Z'} \sim T$, on-shell production of $Z'$ becomes possible and the interaction rate is resonantly enhanced, calculated most easily in the narrow-width approximation for the $Z'$ propagator
\begin{align}
 \frac{1}{(s-M_{Z'}^2)^2 + M_{Z'}^2 \Gamma_{Z'}^2} && \to && \frac{\pi}{M_{Z'} \Gamma_{Z'}} \delta (s - M_{Z'}^2) \,.
\end{align}
This step is equivalent to simply studying the inverse decay $\overline{f} f \to Z'$, as has been done recently in Ref.~\cite{Ahlgren:2013wba} in a similar context.

\begin{figure}
\centering
\includegraphics[width=0.45\textwidth]{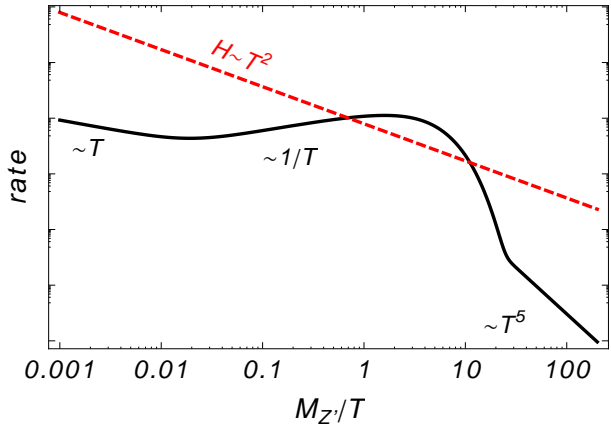}
\caption{Qualitative behavior of the rate $\langle \Gamma (\overline{f} f\leftrightarrow \overline{\nu}_R\nu_R) \rangle $ (black) as a function of temperature. Also shown is the Hubble rate (red, dashed); in this example, the (massless) $\nu_R$ would enter equilibrium around $T \sim M_{Z'}$ and decouple around $T\sim M_{Z'}/10$.}
\label{fig:rate}
\end{figure}

The thermally averaged interaction rate between right-handed neutrinos and massless fermions $f$ via $s$-channel $Z'$ exchange then takes the form
\begin{align}
 &\langle \Gamma (\overline{f} f\leftrightarrow \overline{\nu}_R\nu_R) \rangle  = N_C(f) [Q_{B-L}(f)]^2 \frac{g'^4}{36 \pi^3 \zeta (3)} T\\
 &\quad \times \begin{cases}
         \frac{\pi^4}{144} \,, & x \lesssim \sqrt{\epsilon}\,,\\
          1.15 \frac{ \pi}{8} \frac{M_{Z'}}{\Gamma_{Z'}} \frac{x^3}{e^x -1} \,, & \sqrt{\epsilon} \lesssim  x \lesssim 14 \sqrt{\log \epsilon^{-1}}\,,\\
         \frac{49\pi^8}{2700} x^{-4}\,, & x \gtrsim 14 \sqrt{\log \epsilon^{-1}}\,,
        \end{cases}
 \label{eq:rate}
\end{align}
with $x \equiv M_{Z'}/T$ and $\epsilon \equiv \Gamma_{Z'}/M_{Z'}\ll 1$, illustrated in Fig.~\ref{fig:rate}. Here, $N_C (f)$ [$Q_{B-L}(f)$] denotes the color multiplicity [$B-L$ charge] of fermion $f$.
The rate reduces to the well-known limits for off-shell $Z'$ exchange $\langle \Gamma \rangle \propto (g'/M_{Z'})^4 T^5$ for $M_{Z'}\gg T$ and $\langle \Gamma \rangle \propto g'^4 T$ for $M_{Z'}\ll T$, but is also applicable in the resonant region $M_{Z'}\sim T$.
Due to the thermal distribution of plasma particles, the on-shell production (inverse decay) is feasible over a wide range of temperatures,  dominating the $\nu_R$ interaction rate roughly in the range $\sqrt{\epsilon} \lesssim  x \lesssim 14 \sqrt{\log \epsilon^{-1}}$. In this intermediate regime we find the function given in Eq.~\eqref{eq:rate} that fits well to the numerical results of the narrow-width approximation;\footnote{Assuming Maxwell--Boltzmann statistics for the $\nu_R$ allows for an analytic evaluation of the narrow-width integral and replaces the intermediate function in Eq.~\eqref{eq:rate} by $\pi x^3 K_1 (x)/4\epsilon$.} in particular, it yields the characteristic $\langle \Gamma  \rangle \propto g'^2 M_{Z'}^2/T$ behavior for inverse decays for $x< 1$~\cite{Ahlgren:2013wba}. For $x\gtrsim 1$, the rate is suppressed by a Boltzmann factor $e^{-x}$ as expected.

For the total rate one has to sum over all fermions~$f$ that are in equilibrium at temperature $T$. Following our discussion of $Z'$ branching ratios to light hadrons in Sec.~\ref{sec:model} we only consider the coupling to leptons. We also reintroduce the lepton mass thresholds in our numerical calculations to improve our accuracy.

With the interaction rate at our disposal, we can study $\nu_R$ thermalization; from Fig.~\ref{fig:rate} it is evident that the right-handed neutrinos are either out of equilibrium during radiation domination ($\langle \Gamma  \rangle < H(T)$), or go in \emph{and out} of equilibrium during some epoch. If the $\nu_R$ are in equilibrium and go out before the Universe cools down to about $T(\nu_R) \sim 150$--$\unit[200]{MeV}$~\cite{Anchordoqui:2012qu}, they will contribute to $N_\mathrm{eff}$ in an entropy-suppressed fashion~\cite{Steigman:1979xp, Olive:1980wz}, namely $\Delta N_\mathrm{eff} < 1$, compatible with current data. For masses $M_{Z'} > \unit[10]{GeV}$, the $\nu_R$ decoupling before $T(\nu_R) \sim \unit[150]{MeV}$ then gives the well-known limits of the form $M_{Z'}/g' > \unit[6.7]{TeV}$~\cite{Barger:2003zh}. For $\unit[1]{GeV}< M_{Z'} < \unit[10]{GeV}$, the limits on $g'$ strengthen considerably due to resonant $Z'$ production (see Fig.~\ref{fig:limits}), down to $g'< 6\times 10^{-9}$ at $M_{Z'}= \unit[1]{GeV}$. For lower masses, it becomes impossible to decouple at $T(\nu_R)$, and we have to demand that the (resonant) interaction rate is smaller than $H(T)$ for all temperatures $\unit[1]{MeV}< T < \unit[150]{MeV}$.\footnote{Because of this, only the limits for $M_{Z'}\gtrsim \unit[1]{GeV}$ are sensitive to the precise bound we use for $N_\mathrm{eff}$.} Brushing the resonance peak---sitting at $x\simeq 2.8$---against $H(T)$ then yields the limit $g' \lesssim  10^{-9}\sqrt{M_{Z'}/\unit[100]{MeV}}$ for $\unit[1]{MeV} \lesssim M_{Z'} < \unit[1]{GeV}$. 

For $Z'$ masses below MeV, we demand that the $\nu_R$ come in to thermal equilibrium \emph{after} $T\sim \unit[1]{MeV}$, so $\langle \Gamma  \rangle (\unit[1]{MeV}) < H (\unit[1]{MeV})$. Since the rate goes with $\langle \Gamma  \rangle \propto g'^2 M_{Z'}^2/T$ initially (Fig.~\ref{fig:rate}), the limits are of the form $g' \lesssim  \unit[3\times 10^{-7}]{keV}/M_{Z'}$, as expected from inverse decay~\cite{Ahlgren:2013wba}. (Note that such a light $Z'$ starts to contribute to $N_\mathrm{eff}$ itself, in addition to the $\nu_R$.)
Finally, for $M_{Z'} < \unit[10]{eV}$, on-shell $Z'$ production at BBN temperatures becomes sub-dominant to the off-shell rate $\langle \Gamma  \rangle \propto T$ and the limit becomes independent of $M_{Z'}$: $g' < 2.5\times 10^{-5}$~\cite{Masso:1994ww}.

Overall, we find that BBN gives the strongest constraints on unbroken $B-L$ for $Z'$ masses between $\unit[100]{MeV}$ and $\unit[100]{GeV}$, only briefly surpassed by BaBar (see Fig.~\ref{fig:limits}). From keV to $\unit[100]{MeV}$ they are however less stringent than stellar evolution bounds. Resonant $Z'$ production dominates over the previously used approximations for masses spanning nine orders of magnitude ($\unit[10]{eV} < M_{Z'} \lesssim \unit[10]{GeV}$), and is surely of interest for other models with mediator particles in this range.

We stress again that a thermalization of $\nu_R$ via $Z'$ is not problematic for the baryon-asymmetry mechanism neutrinogenesis~\cite{Dick:1999je}, because the $Z'$ interactions conserve individual particle number and hence would not erase an existing $\nu_R$ asymmetry.

\subsection{Collider}

\begin{figure*}
\centering
\includegraphics[width=0.99\textwidth]{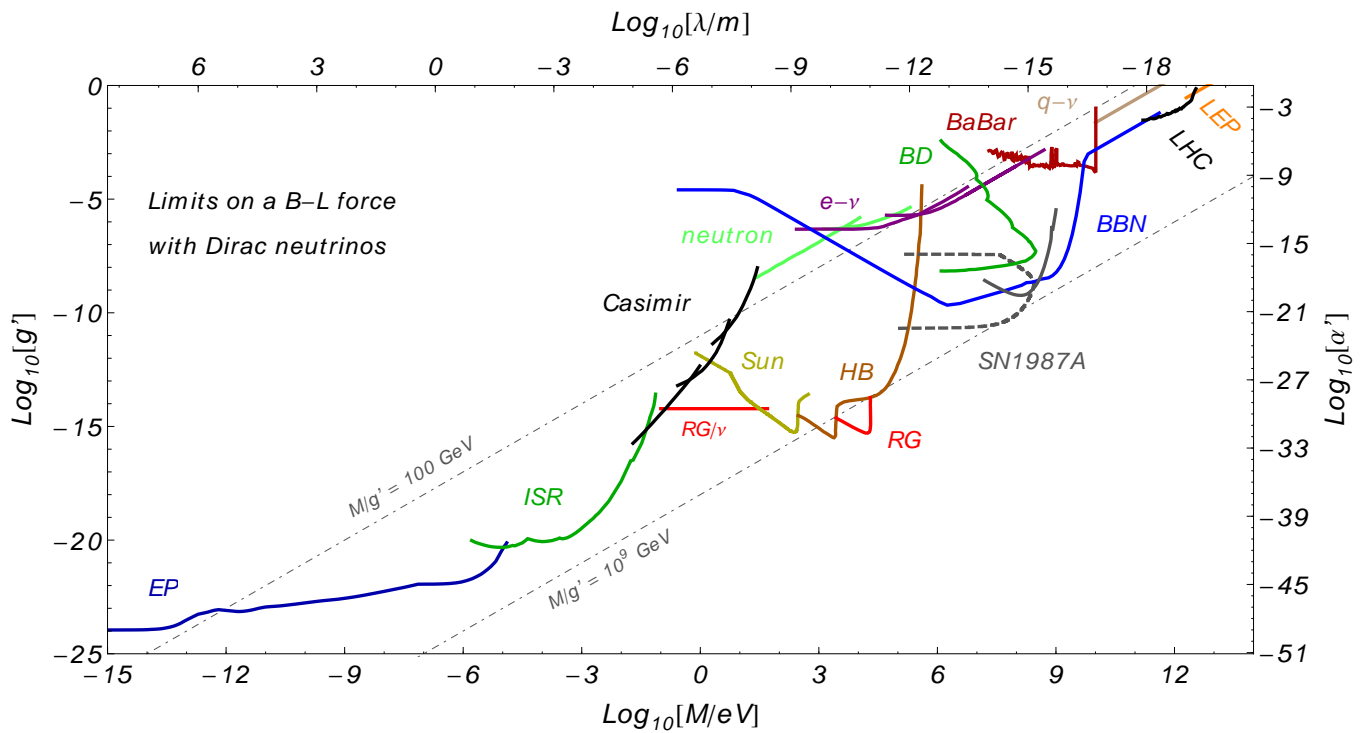}
\caption{Bounds on the $U(1)_{B-L}$ coupling constant $g'$ (fine-structure constant $\alpha = {g'}^2/4\pi$) and $Z'$ St\"uckelberg mass $M_{Z'}$ (corresponding to a range $\lambda = 1/M_{Z'}$) for vanishing kinetic-mixing angle $\chi=0$.
The area on the upper left is excluded by limits from 
tests of the equivalence principle (EP, dark blue)~\cite{Wagner:2012ui, Adelberger:2009zz}, 
gravitational inverse-square law (ISR, green)~\cite{Adelberger:2009zz}, 
Casimir effect (black)~\cite{Bordag:2001qi, Decca:2005qz, Sushkov:2011zz}, 
neutron scattering (light green)~\cite{Pokotilovski:2006up, Nesvizhevsky:2007by},
energy loss via $\nu$ in red giants (RG/$\nu$, red)~\cite{Davidson:2000hf},
energy loss via $Z'$ in the Sun (Sun, yellow), 
horizontal branch stars (HB, orange), and
red giants (RG, red)~\cite{Redondo:2013lna},
SN1987A ($Z'$) (grey, dashed)~\cite{Dent:2012mx},
SN1987A ($\nu_R$) (grey)~\cite{Dreiner:2013mua},
BBN (blue, Sec.~\ref{sec:bbn}),
beam dump searches (BD, green)~\cite{Andreas:2012mt},
neutrino--electron scattering ($e$--$\nu$, purple)~\cite{Harnik:2012ni},
BaBar (dark red)~\cite{Lees:2014xha},
neutrino--quark scattering ($q$--$\nu$, brown)~\cite{Escrihuela:2011cf},
ATLAS (LHC, black)~\cite{Aad:2014cka},
and 
LEP (orange)~\cite{Carena:2004xs, Schael:2013ita}.
Most (non-astrophysical) limits are at $95\%$~C.L., see text for details.
Also shown are diagonals at scales $M_{Z'}/g' = \unit[100]{GeV}$ and $\unit[10^9]{GeV}$ for comparison (grey, dotdashed).}
\label{fig:limits}
\end{figure*}

Direct searches for additional neutral gauge bosons have, of course, been performed at colliders.
The ATLAS experiment at the LHC gives limits on the $B-L$ model that can be used directly~\cite{Aad:2014cka}. (Compared to the ``standard'' $B-L$ model, our unbroken realization has an additional decay channel $Z'\to \nu_R\nu_R$, slightly changing the width. This is however irrelevant for the huge $Z'$ masses considered in the ATLAS analysis.) More specifically, the ATLAS analysis uses dielectron and dimuon channels with $\unit[20]{fb^{-1}}$ at $\unit[8]{TeV}$, constraining $Z'$ masses $0.2$--$\unit[3.5]{TeV}$ at $95\%$~C.L.~(Fig.~\ref{fig:limits}).

Following Ref.~\cite{Carena:2004xs}, we can derive an updated short-range limit of 
\begin{align}
 M_{Z'}/g' > \unit[6.9]{TeV} \, \text{ at } 95 \% \text{ C.L.}
 \label{eq:short-range}
\end{align}
from effective four-lepton operators with the final LEP~2 data~\cite{Schael:2013ita}, valid for $M_{Z'} \gg \unit[200]{GeV}$. Even stronger limits can likely be obtained from updated dedicated global fits to electroweak precision data~\cite{Cacciapaglia:2006pk, Salvioni:2009jp}, but we take Eq.~\eqref{eq:short-range} as the strongest limit on $B-L$ for $Z'$ masses above $\unit[3]{TeV}$.

Not shown in Fig.~\ref{fig:limits} are the LEP constraints for masses close to the $Z$ mass $M_{Z'}\simeq M_Z$, which are stringent but very narrow~\cite{Williams:2011qb}.

\section{Comments}
\label{sec:comments}

We have derived limits on our specific model of unbroken $B-L$ presented in Sec.~\ref{sec:model}, but our results can be translated to other scenarios, too. Since an exhaustive list of related models is infeasible, we stick to the most obvious ones.

\subsection{Broken $B-L$}

The limits of Fig.~\ref{fig:limits} are directly applicable to models with spontaneously broken gauged $U(1)_{B-L}$ and \emph{Dirac} neutrinos, i.e.~scenarios where $B-L$ is broken by $n\neq 2$ units~\cite{Heeck:2013rpa}. The scale $M_{Z'}/g' = n \langle\phi \rangle$ is then directly connected to the $B-L$ breaking vacuum expectation value $\langle\phi \rangle$ of a scalar $\phi$. Even though (non-resonant) Dirac leptogenesis works quite naturally at high $B-L$ breaking scales in this scenario~\cite{Heeck:2013vha}, this does not preclude sub-TeV $Z'$ bosons, because the coupling $g'$ can be small. Seeing as astrophysics is sensitive to scales $M_{Z'}/g' \gtrsim \unit[10^9]{GeV}$ (Fig.~\ref{fig:limits}), even those natural leptogenesis scales are probed (in some select areas at least).

If the $U(1)_{B-L}$ is broken by $n=2$ units, one recovers the more familiar models in which neutrinos are Majorana particles, typically generated by the seesaw mechanism. Both seesaw and thermal leptogenesis hint at high $B-L$ breaking scales, but resonant low-scale models are possible, too. The same argument from above applies, in that even high-scale realizations can have sub-TeV $Z'$ gauge bosons, subject to the constrains from Fig.~\ref{fig:limits}.
In those models, the validity of our derived limits depends only on the mass of the ``heavy'' right-handed neutrinos. If they are far below $M_{Z'}$, the constraints from Fig.~\ref{fig:limits} at said $M_{Z'}$ apply; if all right-handed neutrinos are heavier than the $Z'$, the BBN limits above $M \sim \unit[1]{MeV}$ seize to be valid, but the sub-MeV limits from BBN still apply, because the $Z'$ contributes to $N_\mathrm{eff}$~\cite{Ahlgren:2013wba}. The SN1987A bound from $\nu_R$ emission~\cite{Dreiner:2013mua} (solid grey in Fig.~\ref{fig:limits}) also becomes invalid if all $\nu_R$ are heavy. The other limits do not change qualitatively, because they depend only slightly on the invisible $Z'$ width, and can be translated straightforwardly.

The high scales probed by astrophysics mentioned here are of course not only reminiscent of leptogenesis scales, but also those of reheating and inflation, which can be linked to $Z'$ physics and dark matter (see for example Ref.~\cite{Chu:2013jja}). A discussion goes unfortunately beyond the scope of this letter but certainly deserves attention.

\subsection{Kinetic mixing}

Having ignored kinetic mixing up until now, let us comment on it. Kinetic mixing is allowed in the Lagrangian via $\frac12\sin \chi F^Y_{\mu\nu}F'^{\mu\nu}$, unless some embedding into a Grand Unified Theory is assumed, which is incompatible with our assumption of unbroken $B-L$. Even if zero at some scale, $\chi\neq 0$ will be induced radiatively because we have particles charged under both $U(1)$ groups~\cite{Holdom:1985ag}, generating a coupling of $Z'$ to hypercharge. This makes unbroken $B-L$ a three-parameter model (not counting neutrino masses and mixing), described by $g'$, $\chi$, and $M_{Z'}$. Limits on these parameters have been discussed for $M_{Z'} >\unit[1]{MeV}$ in Ref.~\cite{Williams:2011qb} (albeit without the coupling to light right-handed neutrinos). Let us therefore discuss the low-mass region. Note that our $Z'$ reduces to the familiar hidden photon in the limit $g'\to 0$.

For $M_{Z'}\ll M_Z$, kinetic mixing effectively induces $Z'$ mixing with the photon, so the $Z'$ now couples to a linear combination of $B-L$ and electric charge $Q$. If, for a certain $Z'$ mass, the $B-L$ coupling dominates, we can apply the limits from this letter; if the coupling to electric charge dominates ($|g'| \ll |e \cos \theta_W \sin\chi|$), one can simply take the limits from hidden photons~\cite{Essig:2013lka}. The intermediate regime, where the coupling strengths are of similar order, is more interesting; depending on the relative sign, the limits can either become stronger or cancel each other out. In the most extreme case, the $Z'$ couples to the linear combination $(B-L) - Q$, which vanishes for electrically charged leptons and baryons. This is then a non-chiral force acting exclusively on neutral fermions: neutrons and neutrinos. Quite surprisingly, even this restricted scenario is easily constrained: the fifth force (modified gravity) limits are all still applicable, as are the BBN bounds ($\nu_L\nu_L\leftrightarrow \nu_R\nu_R$ being allowed); stellar evolution can still put constraints on the coupling in the intermediate mass range $\unit{eV} <M_{Z'} <\unit[100]{MeV}$ from the coupling to neutrons, but requires a dedicated reanalysis beyond the scope of this letter. (This is certainly a worthwhile endeavor, as such limits would also be relevant for popular light-mediator models coupled to baryon number.)

Note that the extreme destructive interference of $B-L$ coupling and kinetic mixing, i.e.~the coupling to $(B-L) - Q$, is not only fine-tuned to begin with, it is also unstable with respect to renormalization group evolution. It is thus a convenient limit to derive the most conservative bounds on unbroken $B-L$, but hardly a realistic model.

For higher masses, $M_{Z'}\sim M_Z$, kinetic mixing will rather induce $Z$--$Z'$ mixing, and the $Z'$ inherits a coupling to the weak neutral current~\cite{Williams:2011qb}. The chiral nature of the latter makes a complete cancellation to any particle \`a la $(B-L) - Q$ impossible, and the limits more straightforward to adapt. For $M_{Z'}>M_Z$, a $Z'$ coupled to $B-L$ and hypercharge---corresponding precisely to the case of $U(1)_{B-L}$ plus kinetic mixing---has been discussed in Ref.~\cite{Salvioni:2009mt}.

\section{Conclusion}
\label{sec:conclusion}

Promoting the only ungauged anomaly-free symmetry of the Standard Model, $U(1)_{B-L}$, to a local symmetry requires the introduction of three right-handed neutrinos, which make neutrinos massive. Experimental searches have yet to confirm the prevailing notion amongst theorists that neutrinos are Majorana particles, and hence $B-L$ broken by two units in nature. It is therefore important to point out that there is currently no theoretical or phenomenological argument against an \emph{unbroken} $U(1)_{B-L}$ gauge symmetry, featuring Dirac neutrinos and neutrinogenesis to explain the baryon asymmetry of the Universe. We can even give the new gauge boson $Z'$ a St\"uckelberg mass $M_{Z'}$, a new dimensionful parameter disconnected from other scales. In this letter we collected and updated constraints on the gauge coupling strength $g'$ in the entire testable mass range $M_{Z'} = 0$--$\unit[10^{13}]{eV}$, assuming for simplicity vanishing kinetic mixing with hypercharge.
The excluded scales for a not-too-light $Z'$ range from $M_{Z'}/g' \gtrsim \unit[7]{TeV}$ (BBN and LEP) all the way up to $M_{Z'}/g' \gtrsim \unit[3\times 10^{10}]{GeV}$ (red giants), also applicable to models with broken $B-L$.
We have shown in particular that successful Big Bang nucleosynthesis provides strong constraints in the mass range $\unit[10]{eV}<M_{Z'}<\unit[10]{GeV}$ due to resonant enhancement of the thermalization rate $\overline{f}f\leftrightarrow \overline{\nu}_R\nu_R$, previously unexplored.

\begin{acknowledgments}
The author gratefully thanks Joerg Jaeckel and Joachim Kopp for discussions and comments on the manuscript, and Hye-Sung Lee and Sean Tulin for helpful correspondence.
This work was supported by the IMPRS-PTFS and by the Max Planck Society in the project MANITOP.
\end{acknowledgments} 


\bibliography{bibfile}

\bibliographystyle{utcaps_mod}

\end{document}